\documentclass[10pt,conference]{IEEEtran}
\IEEEoverridecommandlockouts
\usepackage[normalem]{ulem}
\usepackage{algorithm}
\usepackage{braket}
\usepackage[noend]{algpseudocode}
\usepackage{amssymb}
\usepackage{xcolor}
\usepackage{textcomp}
\usepackage{gensymb}

\usepackage[top=1in, bottom=1in, left=1in, right=1in]{geometry}
\usepackage{graphicx,amsmath,amsthm,cite,nicefrac,ifthen,amsfonts}
\usepackage{subfigure}
\usepackage{subcaption}
\usepackage{wrapfig}

\newcommand{\getCurrentSectionNumber}{%
  \ifnum\c@section=0 %
  \thechapter
  \else
  \ifnum\c@subsection=0 %
  \thesection
  \else
  \ifnum\c@subsubsection=0 %
  \thesubsection
  \else
  \thesubsubsection
  \fi
  \fi
  \fi
}

\theoremstyle{definition}

\theoremstyle{definition}

\theoremstyle{definition}

\makeatletter
\def\tagform@#1{\maketag@@@{\bfseries(\ignorespaces#1\unskip\@@italiccorr)}}
\renewcommand{\eqref}[1]{\textup{{\normalfont(\ref{#1}}\normalfont)}}
\makeatother

\begin{document}

\title{A Bi-Objective Routing Framework for Hybrid Terrestrial--Satellite Quantum Networks}
\author{
\IEEEauthorblockN{Yashpreet Khambay, Nitish K. Panigrahy}
\IEEEauthorblockA{Binghamton University, Binghamton, NY, USA \\
Email:\{ykhamba1, npanigrahy\}@binghamton.edu}
}

\maketitle

\begin{abstract}
Hybrid terrestrial-satellite quantum networks combine terrestrial fiber infrastructure with free-space links to enable long-distance entanglement distribution. Entanglement can be routed over different combinations of terrestrial and satellite links, resulting in paths with different entanglement generation rates (EGRs) and fidelities. Existing routing algorithms, however, typically reduce routing to a single-objective problem by optimizing either EGR or fidelity, or by optimizing one while constraining the other, and thus do not explicitly capture the trade-off between the two. In this paper, we present a bi-objective routing framework for hybrid quantum networks that jointly optimizes end-to-end EGR and fidelity. We formulate routing as a Pareto optimization problem and show that it possesses a special mathematical structure: the bottleneck nature of end-to-end EGR and the multiplicative nature of end-to-end fidelity allow the problem to be transformed into a MAXMIN--MINSUM bicriterion path problem. This transformation enables the exact polynomial-time computation of a minimal complete Pareto set using Martins' bicriterion routing algorithm. Our results demonstrate that hybrid networks expose substantially richer Pareto frontiers than terrestrial-only networks and that the proposed framework outperforms representative single-objective routing policies by allowing applications to select paths based on their EGR and fidelity requirements.
\end{abstract}


\section{Introduction}
Quantum networks will enable applications such as quantum key distribution (QKD) \cite{qkdbennett2014quantum}, blind quantum computing (BQC) \cite{barz2012demonstration} and high-precision quantum sensing \cite{komar2014quantum} by distributing entangled states between geographically separated users. Such entanglement distribution can be achieved using two  infrastructures: terrestrial fiber networks \cite{chung2021illinois, bersin2024development} and satellite-based quantum links \cite{yin2017satellite, lu2022micius}.

The two infrastructures, however, exhibit complementary characteristics. Fiber links are generally available for entanglement generation, but suffer exponential transmission loss with distance \cite{pirandola2017fundamental}, requiring quantum repeaters \cite{briegel1998quantum} to establish long-distance entanglement. Satellite links avoid most propagation loss by transmitting through free space and have demonstrated entanglement distribution over more than $1200$ km \cite{yin2017satellite}, but they are available only during visibility windows and remain sensitive to weather conditions. Further, due to the high cost of building optical ground stations \cite{riesing2017portable}, their deployment is limited. These complementary properties have motivated growing interest in \emph{hybrid terrestrial-satellite quantum networks} \cite{bhaskar2021hybrid, harney2022end, shao2025hybrid, yehia2024connecting, gu2025quesat, hu2024dynamic, bakker2024best, shaban2024sparq}, where satellite links establish long-distance entanglement between distant ground stations, while terrestrial fiber networks distribute the resulting entangled states to the end users.

\begin{figure}[htbp]
    \centering
    \includegraphics[width=0.8\columnwidth]{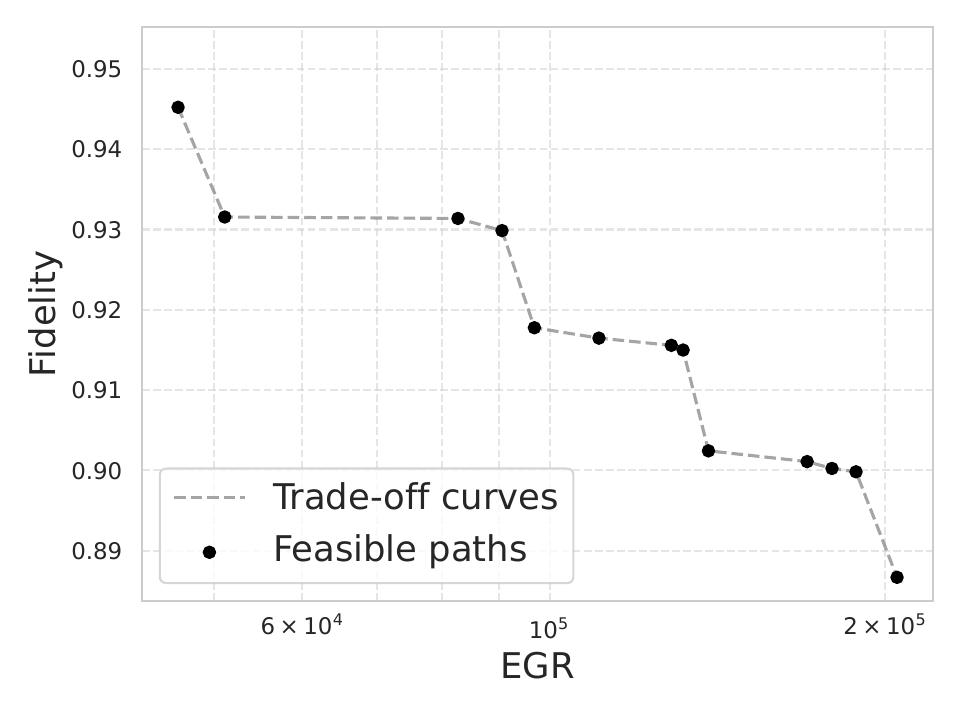} 
    \caption{Trade-off between end-to-end entanglement generation rate (EGR) and fidelity for routing between representative end users in Iselin, NJ and Chicago Ridge, IL. Each point represents a feasible route, while the dashed line highlights the rate-fidelity trade-off.}
    \label{fig:pareto-frontier}
\end{figure}

Recent work \cite{bhaskar2021hybrid, harney2022end, shao2025hybrid, yehia2024connecting} has investigated the  architecture and feasibility of such hybrid networks, with more recent efforts beginning to address networking aspects including topology management and routing \cite{gu2025quesat, hu2024dynamic, bakker2024best, shaban2024sparq}. Nevertheless, efficiently routing entanglement through a dynamic hybrid network remains challenging, as candidate routes often differ substantially in both the entanglement generation rate (EGR) and the quality (fidelity) of the entanglement they can deliver.

Figure \ref{fig:pareto-frontier} illustrates this trade-off for a representative entanglement distribution request between two end-users located in New Jersey and Illinois over a hybrid quantum network. Each point corresponds to a feasible end-to-end route computed under our physical model, accounting for satellite positions, transmission loss and noise, and repeater operations. The figure shows that multiple routing choices are simultaneously optimal: improving the achievable EGR generally comes at the expense of the fidelity of the distributed entanglement. 

Existing routing algorithms typically identify a single path by maximizing one metric (e.g., EGR) or by optimizing one metric subject to a threshold on another (e.g., fidelity). However, no single path may be the best for all applications. For example, consider two quantum sensing applications. Very-long-baseline interferometry \cite{gottesman2012longer} generally favors higher EGR, whereas synchronized atomic clocks \cite{komar2014quantum} require higher-fidelity entangled states. As we demonstrate later (Section~\ref{simulations}), hybrid networks combine terrestrial and satellite-assisted paths to provide a richer set of competitive paths for long-distance entanglement distribution. These paths exhibit distinct rate-fidelity characteristics, creating an opportunity to match each application's requirements to the most appropriate path.

In this paper, we develop a bi-objective routing framework for hybrid quantum networks that jointly considers EGR and fidelity to identify efficient routing paths in polynomial time. Instead of searching for a single best path, we characterize the \emph{Pareto frontier} \cite{martins1984specialclass} of end-to-end paths for each network snapshot. Each point on the frontier represents a feasible routing solution for which the achievable EGR cannot be increased without reducing fidelity and vice versa. 

Although generating the Pareto-optimal routes, in general, is computationally hard \cite{serafini1987some}, we show that the unique bottleneck and multiplicative structure of the proposed routing problem admits an exact polynomial-time solution. The resulting frontier provides a network controller with multiple operating points from which the most appropriate path can be selected according to the application's requirements. 

The main contributions of this paper are as follows: 
\begin{itemize}
 \item We formulate routing in a time-varying hybrid quantum network as a bi-objective optimization problem that jointly optimizes end-to-end EGR and fidelity. Unlike existing routing approaches that optimize a single metric, our formulation explicitly captures the trade-off between EGR and fidelity.

\item We show that the proposed routing problem possesses a special mathematical structure. Specifically, by exploiting the bottleneck nature of end-to-end EGR and transforming the multiplicative fidelity objective into an additive path cost, we reformulate the problem as a MAXMIN-MINSUM bicriterion path problem. This transformation enables an exact polynomial-time computation of a minimal complete set of Pareto-optimal paths using Martins' bicriterion routing algorithm \cite{martins1984specialclass}.
 
\item We conduct a comprehensive evaluation of the proposed framework on realistic terrestrial-only and hybrid terrestrial-satellite quantum networks. Our results quantify the rate-fidelity trade-off, and compare hybrid and terrestrial-only networks, demonstrating the advantages of bi-objective routing over conventional single-objective approaches. 
\end{itemize}
The rest of the paper is organized as follows. Section \ref{rel-lit} reviews related work. Section \ref{QNA} presents the hybrid quantum network architecture. Section \ref{routing} describes the proposed bi-objective routing framework. Section \ref{simulations} evaluates the proposed framework through simulation. Finally, Section \ref{conclusion} concludes the paper.
\section{Related Work}\label{rel-lit}
Quantum network routing research has primarily focused on terrestrial fiber networks. These works demonstrate that optimizing quantum-specific metrics such as EGR and fidelity yields better performance than distance-based metrics such as hop count \cite{van2013pathsdijsktra,gyongyosi2018distancedecentralized}. However, they optimize a single objective or optimize one metric while constraining the other \cite{zhao2022e2efid,caleffi2017optimalegr}. Consequently, they do not explicitly characterize the trade-off between EGR and fidelity.

More recently, hybrid terrestrial--satellite quantum networks have been proposed \cite{shao2025hybrid,bakker2024bestpath,shaban2024sparq,gu2025quesat}. Despite differences in their architectures, including active satellite sources \cite{shao2025hybrid,bakker2024bestpath}, three-tier satellite-air-ground networks \cite{shaban2024sparq}, and passive reflector-based designs \cite{gu2025quesat}, all formulate routing as a single-objective optimization problem.

For example, Shao et al. \cite{shao2025hybrid} propose a hybrid architecture and dynamically switch between terrestrial and satellite links using distance thresholds derived from ground-station density. Similarly, Bakker et al. \cite{bakker2024bestpath} formulate routing as a shortest-path problem by applying Dijkstra's algorithm to a secret-key probability metric that accounts for weather, stray-light radiance and satellite visibility. Shaban et al. \cite{shaban2024sparq} consider a three-tier satellite-air-terrestrial architecture and employ a deep reinforcement learning framework to dynamically select routing paths based on channel transmissivity. Gu et al. \cite{gu2025quesat} place entanglement sources at ground stations, use satellites as passive optical reflectors, and formulate routing as a two-stage mixed-integer linear program. 

Although these approaches differ significantly in their physical architectures and routing algorithms, they all ultimately select a single route by optimizing one routing objective. None explicitly computes the set of Pareto-optimal routing paths that expose the trade-off between EGR and fidelity. While multi-objective optimization has been explored for classical networks under various network configurations \cite{wu2021multiqos,liu2025routinghierarchical}, the unique bottleneck and multiplicative structure of quantum routing objectives presents a fundamentally different optimization problem. In this work, we formulate routing as an exact bi-objective optimization problem and compute the Pareto frontier in polynomial time by transforming it into a MAXMIN-MINSUM bicriterion path selection problem \cite{martins1984specialclass}.

\section{Hybrid Quantum Network Architecture}\label{QNA}
We consider a hybrid quantum network consisting of a terrestrial fiber network and a time-varying LEO satellite network. This section describes the network architecture and defines the link and end-to-end path metrics used for routing.

\subsection{Terrestrial Nodes and Satellites}
The terrestrial layer consists of a set of quantum-capable nodes $\mathcal{V}=\mathcal{V}_g\cup\mathcal{V}_f,$ where $\mathcal{V}_g$ denotes the set of optical ground stations (OGSs) and $\mathcal{V}_f$ denotes the set of terrestrial fiber nodes. In addition to serving as end nodes for entanglement distribution, fiber nodes participate in entanglement swapping. OGSs are additionally equipped with the free-space optical (FSO) hardware required to communicate with satellites. We assume that all terrestrial nodes are equipped with perfect single-photon quantum memories and optical hardware necessary to receive, swap and manipulate entangled states.

The space layer consists of a set $\mathcal{S}$ of LEO satellites. We consider a dual downlink architecture \cite{williams2024scalable, shao2025hybrid} without inter-satellite links. Each satellite is equipped with an entangled-pair source and two optical transmitters. When two OGSs are simultaneously visible to a satellite, the satellite may transmit one photon from an entangled pair to each OGS, thereby creating satellite-assisted entanglement between them.

\subsection{Terrestrial Topology Construction}
We construct the terrestrial topology in two stages. First, a terrestrial edge is established between geographically nearby terrestrial nodes. The resulting proximity-based topology may contain multiple disconnected components. To obtain a globally connected terrestrial network, we iteratively connect the closest pair of nodes belonging to different connected components. Equivalently, this procedure follows an agglomerative single-linkage clustering algorithm (nearest-neighbor technique) \cite{sneath1957applicationagglo}, where the minimum-distance inter-component edge is added at each iteration until the terrestrial graph becomes connected.

Because fiber attenuation causes the EGR to decrease exponentially with distance, we limit the length of each fiber segment to $d_{\mathrm{rep}}$, the maximum distance between two neighboring repeaters. A terrestrial edge whose physical fiber length exceeds $d_{\mathrm{rep}}$ is divided into multiple fiber segments, with repeaters placed between consecutive segments for entanglement swapping. We denote the resulting static set of terrestrial edges by $\mathcal{E}_t$.

\begin{figure}[htbp]
    \centering
    \includegraphics[width=0.8\columnwidth, height = 6.4 cm]{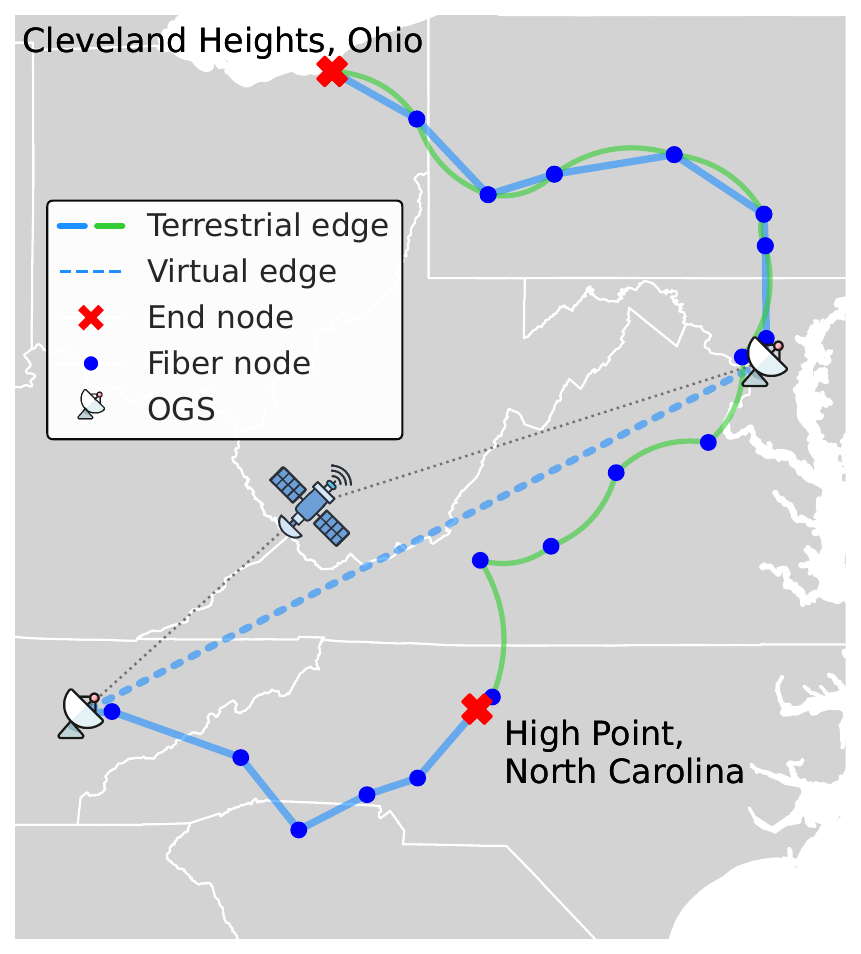} 
    \caption{Example of terrestrial-only and hybrid routes between Cleveland Heights, OH and High Point, NC. The green route uses only terrestrial fiber links, whereas the blue route incorporates a satellite-assisted virtual edge (dashed) between two OGSs.}
    \label{fig:hybrid_fig}
\end{figure}

\subsection{Satellite-Assisted Virtual Links}
Satellite mobility causes the set of available satellite-assisted links to change over time. We therefore divide time into slots of duration $\Delta$ seconds and treat the satellite positions and channel conditions as fixed within each slot. Let $C \subseteq \mathcal{V}_g \times \mathcal{V}_g$ denote the set of OGS pairs that may be served by the satellite layer. At the beginning of slot $t$, we determine which satellites can simultaneously view both OGSs in each pair $(g_1,g_2)\in C$. Assigning a satellite to such a pair creates a transient virtual edge between $g_1$ and $g_2$. The set of virtual edges selected during slot $t$ is
denoted by $\mathcal{E}_s(t)$.

Because the number of simultaneously available satellite transmitters and OGS receivers is limited, not every geometrically feasible virtual edge can be activated. We employ a greedy scheduling procedure subject to the following resource constraints:
\begin{enumerate}
    \item each satellite is assigned to at most one OGS pair during a time slot; and
    \item each OGS pair is served by at most one satellite during that time slot.
\end{enumerate}
The scheduling procedure first considers OGS pairs that are visible to only one satellite. It then prioritizes the remaining pairs in descending order of geographic separation, thereby using the satellite layer to bridge long terrestrial distances. When multiple satellites can serve the same pair, the satellite providing the largest EGR is selected.

A satellite-assisted edge is virtual in the sense that the two OGSs are not connected by a direct physical channel. Instead, the satellite generates an entangled photon pair and transmits one photon to each OGS through two independent free-space downlinks.

\subsection{Time-Varying Multigraph Representation}
We model the resulting hybrid network during slot $t$ as the
time-varying multigraph $\mathcal{G
}(t)=\bigl(\mathcal{V},\mathcal{E}(t)\bigr),$ where $\mathcal{E}(t)=\mathcal{E}_t\cup\mathcal{E}_s(t).$ The terrestrial edge set $\mathcal{E}_t$ remains fixed, whereas the satellite-assisted edge set $\mathcal{E}_s(t)$ is recomputed as satellite
positions and channel conditions change. We use a multigraph because the same OGS pair may be connected simultaneously by a terrestrial fiber edge and by a satellite-assisted virtual edge. These parallel edges correspond to physically distinct entanglement distribution mechanisms and may provide different EGRs and fidelities.

Figure~\ref{fig:hybrid_fig} illustrates the resulting architecture. The terrestrial layer provides fiber connectivity, while the satellite layer dynamically introduces long-distance virtual edges between selected OGS pairs. End-to-end entanglement may therefore be distributed either through a terrestrial-only route or through a hybrid route that traverses both terrestrial fiber edges and satellite-assisted virtual edges.

\subsection{Entanglement Sources and Channel Model} 
We assume that entanglement is generated over every elementary edge by a spontaneous-parametric-down-conversion (SPDC) source \cite{kok2000postselected}. For a terrestrial
fiber edge, the source is placed between the two endpoints of the
edge. For a satellite-assisted edge, the source is located
onboard the satellite and its two output photons are transmitted to the corresponding OGSs. We model an SPDC-based dual-rail polarization source whose output, truncated to the vacuum, single-pair, and two-pair components:
\begin{align}
    \ket{\psi^\pm}
    =&N_0\Bigg[
    \sqrt{p(0)}\ket{0,0;0,0} \nonumber\\
    &+\sqrt{\frac{p(1)}{2}}
    \left(
    \ket{1,0;0,1}
    \pm
    \ket{0,1;1,0}
    \right) \nonumber\\
    &+\sqrt{\frac{p(2)}{3}}
    \left(
    \ket{2,0;0,2}
    \pm
    \ket{1,1;1,1}
    +
    \ket{0,2;2,0}
    \right)
    \Bigg],
    \label{eq:spdc_state}
\end{align}
where $N_0$ is a normalization constant and
\begin{equation}
    p(n)=(n+1)\frac{N_S^n}{(N_S+1)^{n+2}}.
    \label{eq:spdc_distribution}
\end{equation}
Here, $N_S$ is the mean photon number per mode, which is controlled by the source pump power. Increasing $N_S$ increases the probability of generating an entangled pair, but also increases the probability of undesired multi-pair emissions. The source therefore exhibits an inherent rate-fidelity trade-off.

For a fiber edge, the channel transmissivity decreases exponentially with the fiber length. For a satellite-assisted edge, the losses grow more slowly, roughly quadratically with distance. We consider two principal sources of noise. First, multi-pair emissions that may produce detection events that cannot be distinguished from single-pair events after photon loss. Second, detector dark counts and background photons that may produce spurious detection events at the receivers. These effects reduce the fidelity of the successfully distributed entangled state. Background-photon noise may vary over time, particularly between daytime and nighttime operating conditions. A more detailed characterization of the source, the loss and noise model can be found in \cite{Dhara22}.

\subsection{Link Metrics}
\label{sec:link_metrics}
Each edge $e\in\mathcal{E}(t)$ is characterized by the pair $(R_e,F_e)$, where the link EGR $R_e$ is the rate at which entangled pairs are successfully distributed over edge $e$. It depends on the source repetition rate, and channel transmissivity. The entangled state produced by the SPDC source and distributed over an edge is generally an arbitrary two qubit state. For analytical tractability, we approximate it by a \emph{Werner state} with the same fidelity $F_e.$ Its Werner parameter
$W_e$ is related to the fidelity $F_e$ by $W_e=(4F_e-1)/3,$ where $W_e\in (0,1]$ and $F_e\in (1/4,1]$. 
\subsection{End-to-End Path Metrics}
\label{sec:path_metrics}

Consider a path $p=(e_1,\ldots,e_k)$ between a pair of end-users $(s,d)\in\mathcal{V}\times\mathcal{V}$, where each $e_i\in\mathcal{E}(t)$. Intermediate nodes perform entanglement swapping to extend entanglement across edges. We assume deterministic swapping operations, and that quantum memories can store generated entanglement until the neighboring links are ready. Under this model, the end-to-end EGR is determined by the  link with the smallest EGR in the path. Hence, the EGR for a path $p$ is given by $R_p=\min_{e\in p}R_e.$
   
Under entanglement swapping, the Werner parameter of the
resulting end-to-end state is the product of the individual link-level Werner parameters. Therefore, the fidelity for path p is given by $F_p= 3(\prod_{e\in p}W_e+1)/4.$

\section{Bi-Objective Routing Framework}\label{routing}
For a fixed network state $\mathcal{G}(t)$ and a pair of end users $(s,d)\in\mathcal{V}\times\mathcal{V}$ requesting entanglement, let $\mathcal{P}_{sd}$ denote the
set of all feasible paths between them. Each path
$p\in\mathcal{P}_{sd}$ is characterized by its end-to-end EGR $R_p$ and fidelity $F_p$, defined in Section~\ref{sec:path_metrics}.

Quantum applications require both a sufficient EGR and sufficiently high fidelity. These objectives are fundamentally conflicting in hybrid networks. Paths providing higher
EGR often traverse terrestrial edges with fewer satellite-assisted segments, whereas paths preserving higher fidelity may not use long repeater chains, instead use satellite-assisted edges. Consequently, no single path simultaneously optimizes both objectives.

We therefore formulate routing as the following bi-objective
optimization problem:
\begin{equation}
    \max_{p\in\mathcal{P}_{sd}}
    \left(R_p,F_p\right).
    \label{eq:biobj}
\end{equation}

Since the two objectives are conflicting, the solution of
(\ref{eq:biobj}) is generally not a single path, but a set of
Pareto-optimal paths. A path $p\in\mathcal{P}_{sd}$ is Pareto optimal if there exists no other path $q\in\mathcal{P}_{sd}$ satisfying $R_q \ge R_p$, and $F_q \ge F_p,$ with at least one strict inequality. Collectively, these paths define the Pareto frontier, where each point represents the best achievable trade-off between end-to-end EGR and fidelity.

\subsection{Transformation to a MAXMIN-MINSUM Routing Problem}
Although multi-objective shortest path problems are generally NP-hard \cite{serafini1987some}, the proposed routing formulation possesses a special mathematical structure that admits an exact polynomial-time solution.

The first objective has a bottleneck form since $R_p=\min_{e\in p}R_e,$ where $R_e$ denotes the EGR of edge $e$. Thus, maximizing the end-to-end EGR is a MAXMIN objective. The fidelity objective is multiplicative: $F_p=(3\prod_{e\in p}W_e+1)/4,$ where $W_e$ is the Werner parameter of edge $e$. Since $(3x+1)/4$ is strictly increasing, maximizing $F_p$ is equivalent to maximizing $\prod_{e\in p}W_e.$ Define the transformed edge cost
\begin{equation}
    C_e=-\log W_e.
    \label{eq:edge_cost}
\end{equation}
Because $0<W_e\le1$, we have $C_e\ge0$. Furthermore,
\begin{align}
    \arg\max_{p\in\mathcal{P}_{sd}}
    \prod_{e\in p}W_e
    =
    \arg\min_{p\in\mathcal{P}_{sd}}
    \sum_{e\in p}C_e.
\end{align}

Hence, maximizing end-to-end fidelity is exactly equivalent to
minimizing an additive path cost. Consequently, the proposed routing problem can be written as
\begin{equation}
    \left(\max_{p\in\mathcal{P}_{sd}} \min_{e\in p}R_e, \min_{p\in\mathcal{P}_{sd}} \sum_{e\in p}C_e,\, \right),
    \label{eq:minsummaxmin}
\end{equation}
which is a bicriterion routing problem consisting of one bottleneck (\textsc{MAXMIN}) objective and one additive (\textsc{MINSUM}) objective.

\subsection{Exact Computation of Pareto Frontier} \label{sec:exact-frontier}
Martins~\cite{martins1984specialclass} showed that bicriterion routing problems comprising one bottleneck objective and one additive objective can be solved exactly in polynomial time by iteratively solving the additive shortest-path problem while progressively eliminating bottleneck edges.

Our transformed routing problem satisfies these structural
requirements exactly. At each iteration, we first compute the
minimum-cost path with respect to the transformed fidelity costs (\ref{eq:edge_cost}) using Dijkstra's algorithm. Let $R_p$ denote the bottleneck EGR of the selected path. Since every path containing an edge with EGR not exceeding $R_p$ cannot produce a strictly larger bottleneck rate, all such edges are removed from the graph before the next iteration. The algorithm then repeats on the reduced graph until the
pair of end nodes requesting entanglement becomes disconnected.

This iterative pruning progressively increases the bottleneck EGR while computing the highest-fidelity path for each attainable bottleneck value. The resulting sequence of paths contains one representative for every distinct Pareto point, thereby constructing a minimal complete Pareto set.

\subsection{Complexity Analysis}

The bottleneck EGR of any path must equal the EGR of at least one edge on that path. Consequently, there are at most $|\mathcal{E}(t)|$ distinct bottleneck edges. Since each iteration removes all edges having the current bottleneck value, the algorithm performs at most $|\mathcal{E}(t)|$ iterations. Each iteration requires one shortest-path computation over nonnegative edge costs. Using Dijkstra's algorithm with a binary heap, the overall complexity is therefore $O\!(|\mathcal{E}(t)|(|\mathcal{E}(t)|+|\mathcal{V}|)\log|\mathcal{V}|),$
which is polynomial in the size of the network.

\subsection{Path Selection from the Pareto Frontier}\label{sub:pareto-path-selection}
The proposed routing framework computes the complete Pareto frontier, which exposes all efficient routing alternatives for a given end-user pair. In practice, however, only a single path is used to distribute entanglement. The final route therefore depends on the objective of the application requesting entanglement.

Let $\mathcal{P}_{sd}^{\star}$ denote the Pareto-optimal path set computed in Section \ref{sec:exact-frontier}. Given an application-specific utility function $U(R_p,F_p)$, the network controller selects the path
\begin{equation}
    p^{\star}
    =
    \arg\max_{p\in\mathcal{P}_{sd}^{\star}}
    U(R_p,F_p).
    \label{eq:utility_selection}
\end{equation}

This formulation accommodates the requirements of various quantum applications through different utility functions. We consider three representative utilities \cite{vardoyan2023quantumqnum}: 

(i) \emph{Secret key rate (SKR):} Applications based on quantum key distribution \cite{qkdbennett2014quantum} may select the Pareto-optimal path that maximizes the achievable secret key rate. When the entangled states are Werner states, the secret key rate achieved for a given path $p$ is
\begin{align}\label{eq:skr_raw}
     U_p^\text{SKR} =  R_p \max\left\{0, \bigg( 1 - 2 h\bigg( \frac{2}{3} - \frac{2}{3}F_p\bigg)\bigg)\right\},
\end{align}
\noindent where $h$ is the binary entropy function. 

(ii) \emph{Entanglement Negativity:} Negativity captures the amount of  entanglement for bi-partite states with utility \cite{vardoyan2023quantumqnum}
\begin{align}
        U_p^\text{NEG} =   R_p \max\left\{0, \bigg(F_p-\frac{1}{2}\bigg)\right\}.\label{eq:neg}
\end{align}
(iii) \emph{Distillable entanglement:} Similarly, applications requiring
    high-quality entanglement may select the path maximizing the achievable distillable entanglement \cite{vardoyan2023quantumqnum}
\begin{align}
        D_p^\text{DE} &=  \bigg(1+F_p\log F_p + (1-F_p) \log ((1-F_p)/3)\bigg),\nonumber\\
        U_p^\text{DE} &= R_p\max\left\{0, D_p^\text{DE}\right\} .\label{eq:dst}
\end{align}
Since the minimal complete Pareto-optimal paths are computed once, supporting different applications requires only evaluating the corresponding utility over the Pareto frontier rather than rerunning the routing algorithm.

When an application-specific utility is unavailable, the controller may instead select a path that balances EGR and fidelity. Let $\bar{R}_p = (R_p-R_{\min})/(R_{\max} - R_{\min}),$ and $\bar{F}_p = (F_p-F_{\min})/(F_{\max} - F_{\min}).$ Here, $R_{\max} = \max_{q \in \mathcal{P}_{sd}^{\star}} R_q,$ and $F_{\max} = \max_{q \in \mathcal{P}_{sd}^{\star}} F_q.$ Similarly, $R_{\min} = \min_{q \in \mathcal{P}_{sd}^{\star}} R_q,$ and $F_{\min} = \min_{q \in \mathcal{P}_{sd}^{\star}} F_q.$ \emph{The utopia point} \cite{marler2004surveyweightsumandcompromise}, corresponding to the hypothetical point that simultaneously achieves the maximum EGR and maximum fidelity, is then used as the reference. Since this point is generally unattainable, the controller selects the Pareto-optimal path closest to it. 
Under the weighted utopia point method, with the given weights $\omega_R$ and
$\omega_F$, where $\omega_R,\omega_F \geq 0$ and
$\omega_R+\omega_F=1$, the controller selects
\begin{equation}
    p^{\star}
    =
    \arg\min_{p \in \mathcal{P}_{sd}^{\star}}
    \left[
        \omega_R(1-\bar{R}_p)^2
        +
        \omega_F(1-\bar{F}_p)^2
    \right].
\end{equation}
The weights allow the controller to adjust the relative importance of EGR and fidelity.

\section{Performance Evaluation}\label{simulations}
In this section, we evaluate the proposed Pareto Routing (Pareto) framework on a hybrid terrestrial-satellite quantum network and compare its performance with several routing baselines. We first describe the network setup and simulation parameters. 

\begin{figure}[!t]
    \centering
    \includegraphics[width=\columnwidth]{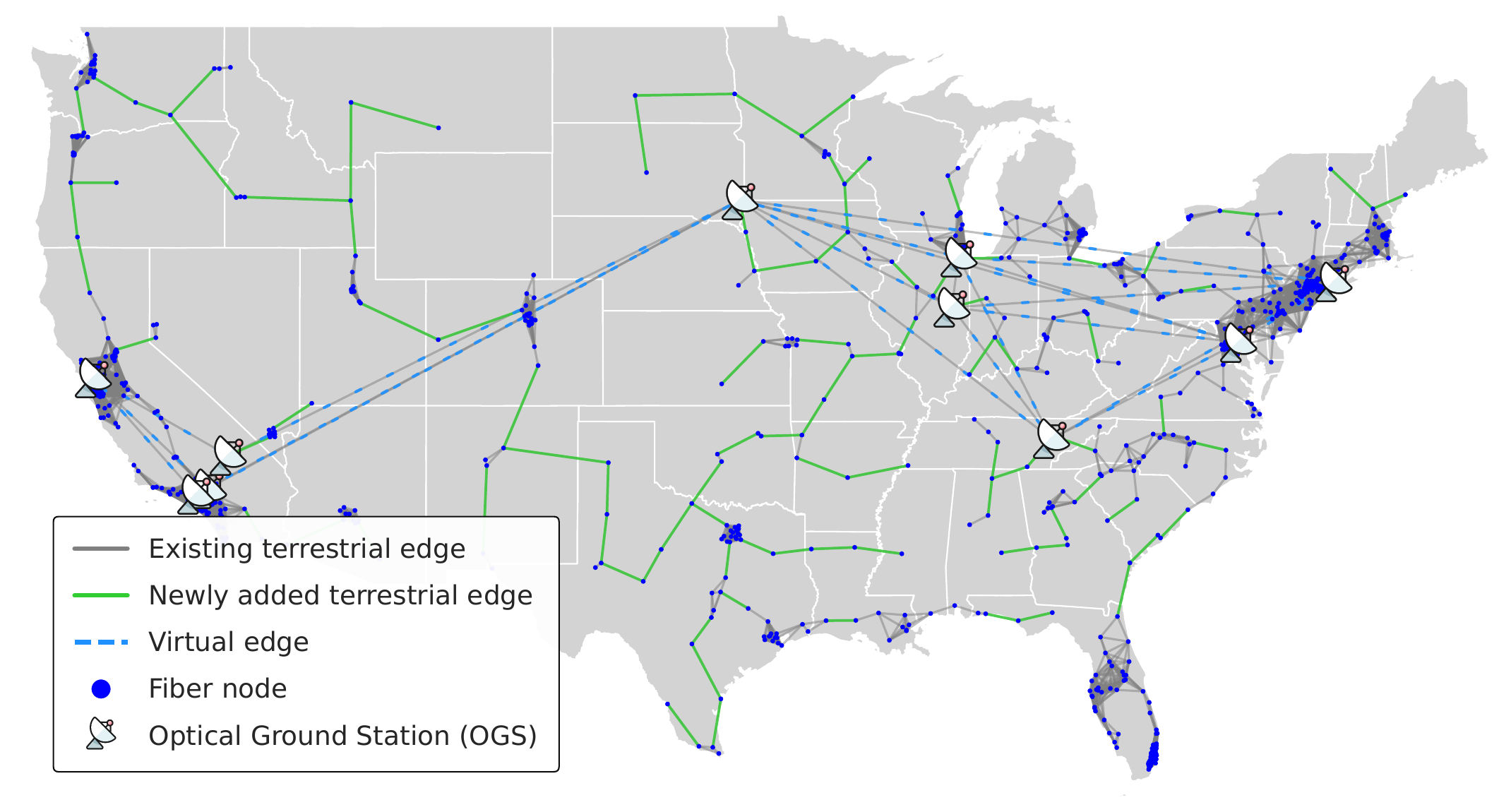} 
    \caption{Simulated hybrid quantum network topology over the United States. The network integrates 1000 fiber end-nodes at the top population centers with 10 strategically selected Optical Ground Stations (OGS). Solid lines (green and gray) represent static terrestrial edges, while dashed lines (blue) indicate satellite-assisted virtual edges for a network snapshot.}
    \label{fig:US_topology}
\end{figure}

\subsection{Simulation Setup} \label{sec:simulation setup}
We evaluate the proposed routing framework using a discrete-time simulator developed in Python 3, using third-party packages like PyEphem, Matplotlib, Networkx, NumPy, and global-land-mask.

\emph{Terrestrial Network.} We simulated the terrestrial network on continental United States (US) with $|\mathcal{V}| = 1010$ terrestrial nodes. To accurately reflect the distribution of realistic user demand, $|\mathcal{V}_f| = 1000$ fiber end-nodes were selected from the top population-centers from the dataset in \cite{Youderain2021worldcitiesdata}. Additionally, $|\mathcal{V}_g| = 10$ OGSs were strategically selected based on existing or planned research infrastructure (see Figure~\ref{fig:US_topology}).

\emph{Satellite Network.} We use the Starlink LEO constellation to represent the satellite layer. We use publicly available two-line element (TLE) data to track the positions of $3,980$ satellites distributed across four orbital inclinations (43$^\circ$, 53$^\circ$, 70$^\circ$, and 97$^\circ$). A satellite is considered visible to an OGS only if its elevation angle exceeds $20^\circ$. At each time slot, we apply the greedy scheduling algorithm described in Section~\ref{QNA} to establish satellite-assisted virtual edges. Specifically, a virtual edge $e_{(g_1,g_2)}\in\mathcal{E}_s(t)$ is created between any OGS pair $(g_1,g_2)$ that simultaneously observes the same satellite, subject to the satellite assignment constraints described in Section ~\ref{QNA}. The satellite-assisted edge set $\mathcal{E}_s(t)$ varies over time as satellite visibility changes.

\emph{Comparison routing policies.}
We compare the proposed Pareto Routing (Pareto) policy with three representative routing policies. Unless stated otherwise, we select a path from the Pareto frontier using the utopia-point method described in Section \ref{sub:pareto-path-selection}.

\begin{itemize}
    \item \emph{Maximum-Fidelity} (Max-Fidelity): Selects the path with the highest end-to-end fidelity, without considering its EGR.

    \item \emph{Fidelity-Constrained Maximum EGR} (FC-MaxEGR):
    Selects the path with the highest EGR among those satisfying a minimum fidelity requirement. Unlike Max-Fidelity, this policy considers both EGR and fidelity, but returns a single path determined by the chosen fidelity threshold. Unless otherwise stated, we set this threshold to $0.5$.

    \item \emph{Distance-Based} (Distance-Based): Uses the geographic distance between the end users to decide whether to use terrestrial or satellite-assisted routing~\cite{shao2025hybrid}. Requests below a distance threshold are routed over the terrestrial network, while requests above the threshold can use satellite-assisted virtual links. We describe the selection of this threshold in the following subsection.
\end{itemize}

\emph{Parameter Selection.} The proposed routing framework introduces two system parameters that govern the construction of the terrestrial topology and the operation of the distance-based routing heuristic.

\begin{figure}[ht]
    \centering
    \hspace{-.3cm}
    \subfigure[]{\includegraphics[width=0.25\textwidth, height=3.2cm]{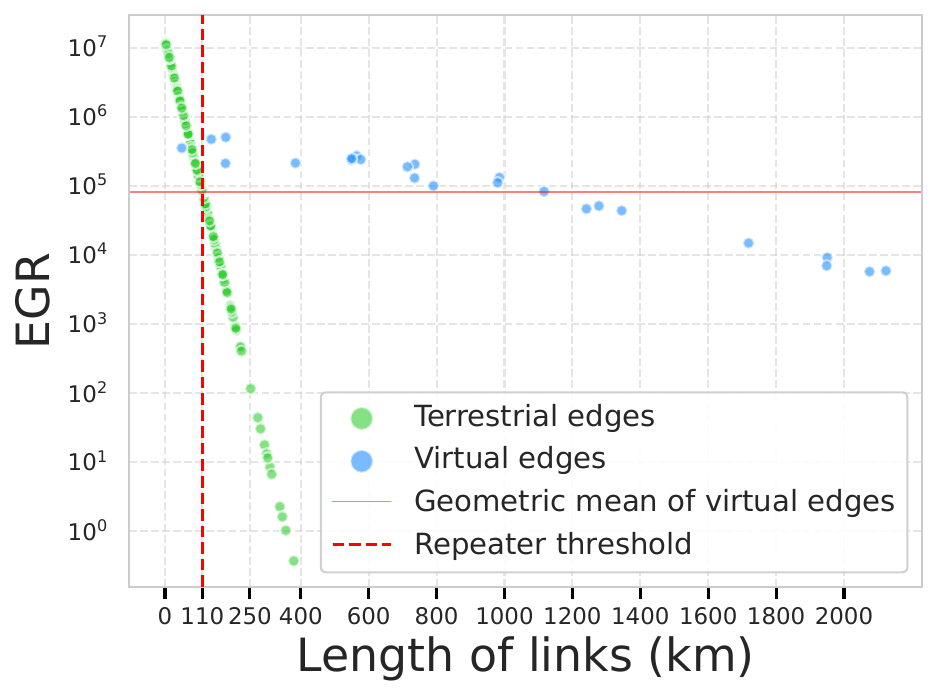}} 
    \hspace{-.3cm}
    \subfigure[]{\includegraphics[width=0.25\textwidth, height=3.2cm]{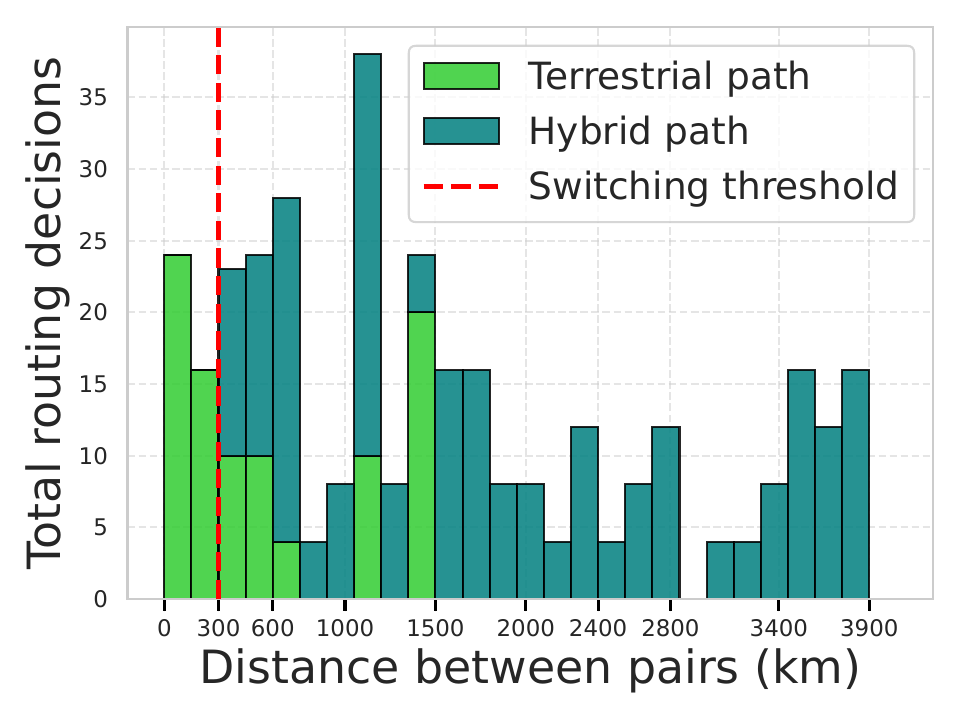}}
    \caption{Selection of (a) the repeater spacing $d_{\mathrm{rep}}$, and (b) the terrestrial-satellite switching distance $d_{\mathrm{switch}}$ used in the baseline distance-based routing.}
    \label{fig:params}
\end{figure}

Figure~\ref{fig:params}(a) shows the achievable EGR for elementary fiber edges as a function of edge length. Beyond approximately $d_{\mathrm{rep}}$, the EGR decreases rapidly, dropping below the geometric mean of virtual edge rate ($\approx10^5$). We limit the length of every elementary fiber edge to $d_{\mathrm{rep}} = 110$ km, inserting intermediate repeaters whenever a fiber connection exceeds this distance. This ensures that entanglement is generated only over physically viable elementary links while longer end-to-end connections are established through entanglement swapping.

Furthermore, we establish direct terrestrial connections between neighboring nodes within the geographical distance threshold $d_{rep} = 110$ km. Initially, this resulted in $94$ disconnected components in the terrestrial topology. Subsequently, to establish a globally connected network, we applied an agglomerative single-linkage clustering algorithm (or nearest-neighbor technique) \cite{sneath1957applicationagglo}. As a result, the total number of static terrestrial edges increased to $|\mathcal{E}_t| =28777$, with the longest edge of length $379$ km connecting a disconnected component and requiring $3$ intermediate repeaters. Figure~\ref{fig:US_topology} shows the resulting topology.

The distance-based routing policy uses terrestrial fiber for nearby end-user pairs and satellite-assisted virtual links for sufficiently separated pairs. To determine this switching threshold $d_{\mathrm{switch}}$, we first apply the remaining routing policies to all end-user pairs and record whether their selected path contains a satellite-assisted virtual link. Figure~\ref{fig:params}(b) shows the fraction of routing decisions utilizing the satellite layer as a function of the geographic separation between end users. Satellite-assisted routes become increasingly preferred beyond $d_{\mathrm{switch}}$, reflecting the point at which the long-distance advantage of free-space transmission outweighs the attenuation incurred over terrestrial fiber. We therefore
use this value as the switching threshold for the Distance-Based routing policy throughout the remainder of the evaluation.


\begin{figure}[!t]
    \centering
    \includegraphics[width=\columnwidth]{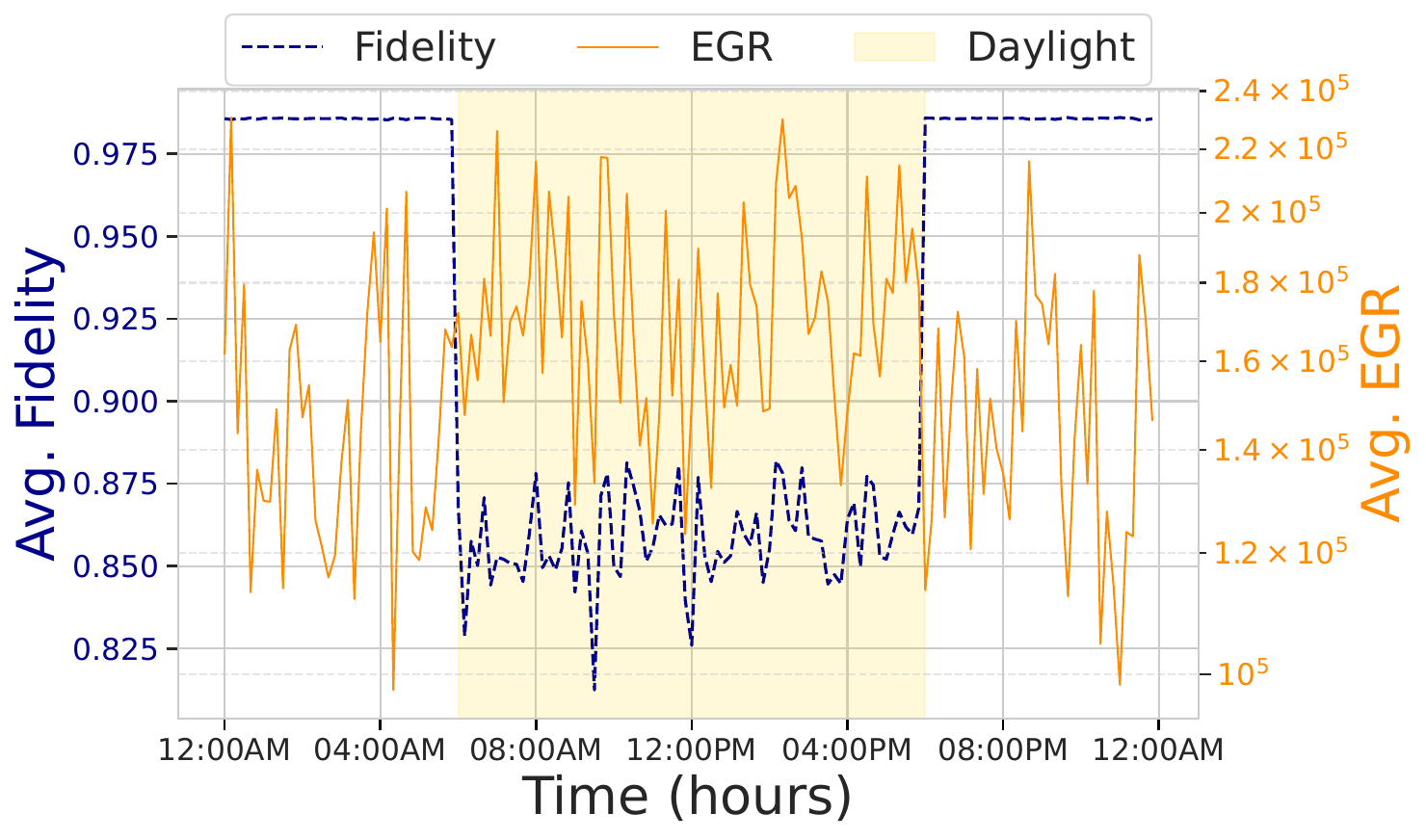} 
    \caption{Temporal characteristics of the satellite-assisted virtual edge set $\mathcal{E}_s(t)$ over a 24-hour period. Daytime solar background noise reduces the average fidelity of satellite-assisted virtual links, whereas nighttime conditions provide consistently higher-quality links.}
    \label{fig:virtual_stats}
\end{figure}

\emph{Channel Loss and Noise.} We set the attenuation coefficient $\beta = 0.2 \text{ dB/km}$ for terrestrial fiber edges and assumed an atmospheric thickness of 5 km above ground level for FSO transmissivity calculations.

We ran the simulation for 24 hours to account for dynamic environmental noise in free space channels. We divided each day into Daytime ($6:00$ AM to $6:00$ PM) and Nighttime ($6:00$ PM to $6:00$ AM) periods. The background noise or detector dark click probability for the space layer was set to a higher value ($P_d = 3 \times 10^{-3}$) for the day, primarily due to background photons emitted by the sun and a lower value ( $P_d = 3 \times 10^{-6}$) for the night.

\subsection{Virtual Link Characteristics}
We analyze the characteristics of the satellite assisted virtual links $\mathcal{E}_s(t)$, over a 24-hour period. Figure~\ref{fig:virtual_stats} shows that at night, FSO links maintain nearly constant high average fidelity ($\approx 0.97$) and rate ($> 10^5$), whereas during daytime hours, we observe a drop in average fidelity by $\approx 18\%$ and an increase in end-to-end EGR. This degradation can become more severe for paths containing multiple satellite-assisted
links, limiting the ability of the satellite layer alone to provide continuous service throughout the day.

\begin{figure*}[htbp]
    \centering
       
    \subfigure[]{\includegraphics[width=0.49\textwidth]{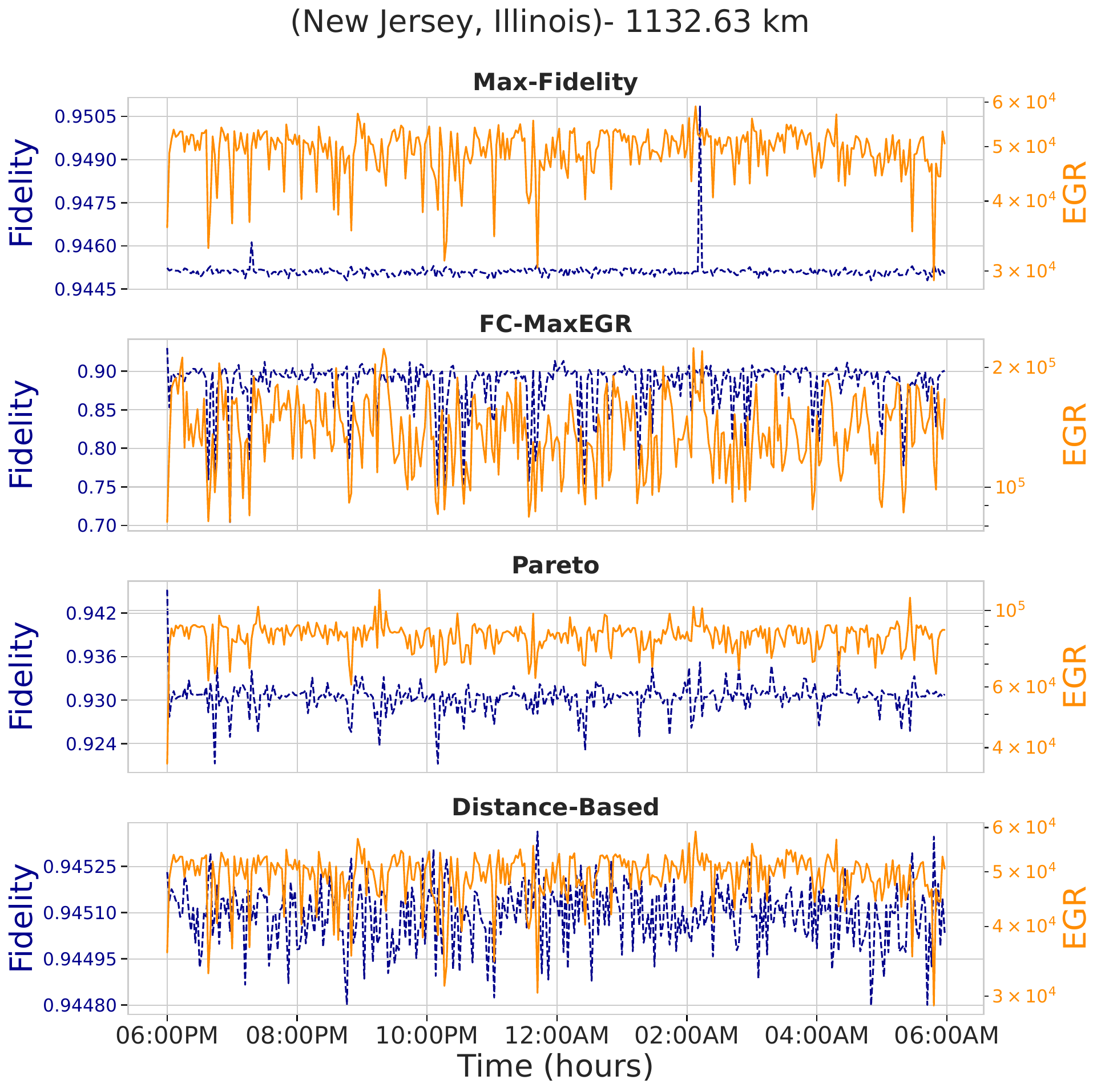}\label{fig:d}}
    \hfil
    \subfigure[]{\includegraphics[width=0.49\textwidth]{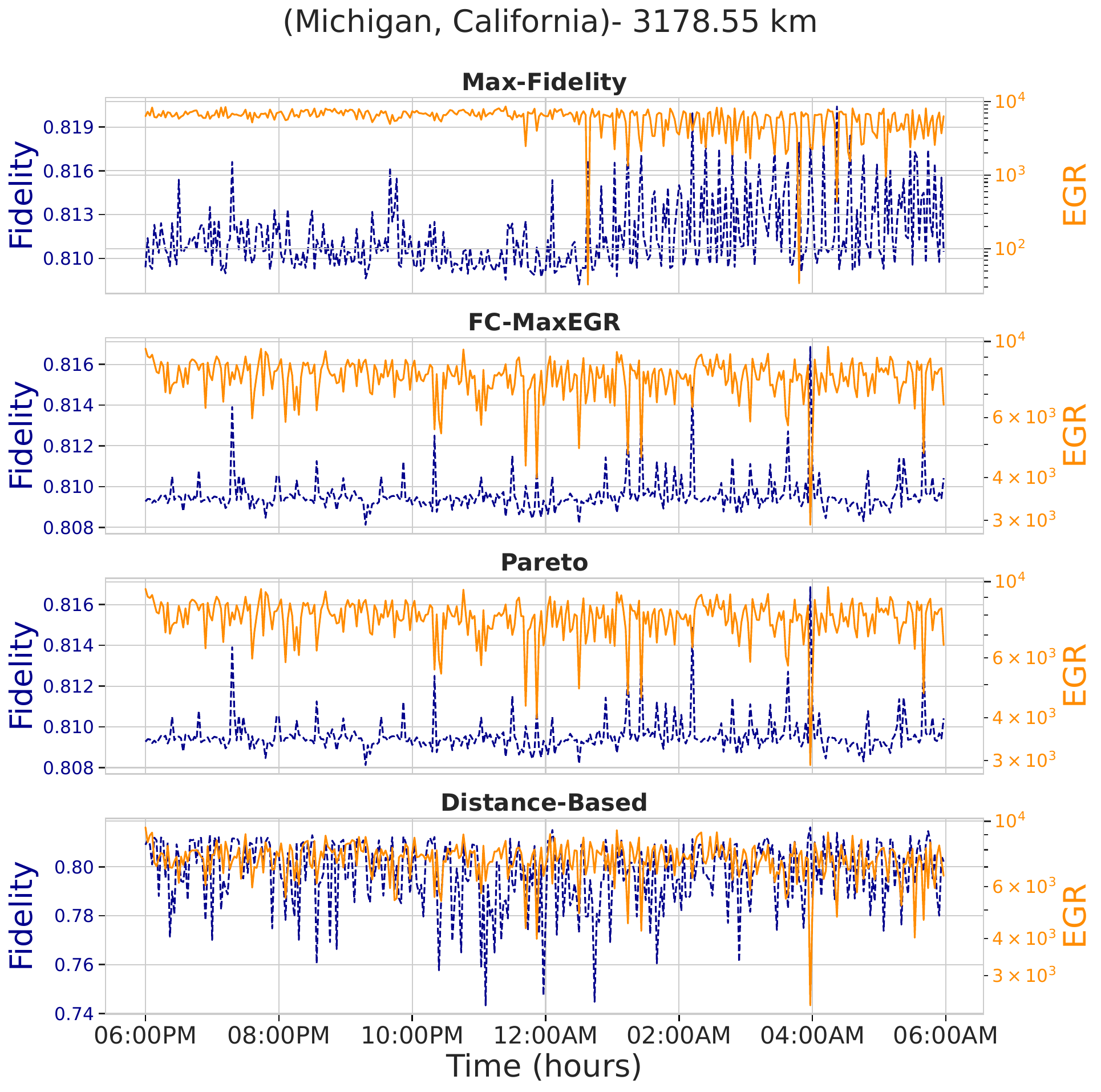}\label{fig:e}}    
    
    \caption{Temporal evolution of end-to-end fidelity and entanglement generation rate (EGR) over a 12-hour period for two representative end-user pairs. Results compare the proposed bi-objective routing policy (Pareto) against the Max-Fidelity, FC-MaxEGR, and Distance-Based baselines. (a) Iselin, NJ to Chicago Ridge, IL (Medium distance). (b) Ann Arbor, MI to North Highlands, CA (Long distance).}
    \label{fig:time_series}
\end{figure*}

\subsection{Algorithm Evaluation}
To evaluate the temporal performance of routing algorithms, we tracked the optimal paths established between 100 randomly selected source-destination pairs $(s,d) \in \mathcal{V}$ over a 24-hour period at 10-second intervals with 8640 total time slots. To better understand the effect of source-destination separation during routing, we divided these pairs into three categories based on their geographical distances, namely short-range ($\le 1000$km, $30$ pairs), medium-range ($>1000$km and $\le 2000$ km, $34$ pairs), and long-range ($>2000$km, $36$ pairs). For each pair and time slot, we recorded the selected path, its end-to-end EGR, and its fidelity.

We first compare the temporal behavior of the routing policies for two representative end-user pairs: a medium-distance pair between Iselin, New Jersey, and Chicago Ridge, Illinois ($1132.63$km) and a long-distance pair between Ann Arbor, Michigan and North Highlands, California ($3178.55$km). Figure~\ref{fig:time_series} reports the end-to-end EGR and fidelity achieved by each policy over a 12-hour period. 

For the medium-distance pair, Max-Fidelity maintains high fidelity but at a relatively low EGR, while FC-MaxEGR achieves a much higher EGR with lower and more variable fidelity. Pareto Routing selects paths between these two extremes, achieving higher EGR than Max-Fidelity while maintaining higher fidelity than FC-MaxEGR.

For the long-distance pair, all policies operate at lower fidelities. Max-Fidelity maintains the highest fidelity but has a low EGR. Pareto Routing and FC-MaxEGR select similar paths under the chosen fidelity threshold, hence their performance appears almost identical with respect to EGR and fidelity. Distance-Based routing shows larger variations in both EGR and fidelity.

\begin{figure}[!ht]
    \centering
    \hspace{-.3cm}
    \subfigure[]{\includegraphics[width=0.25\textwidth, height=3.2cm]{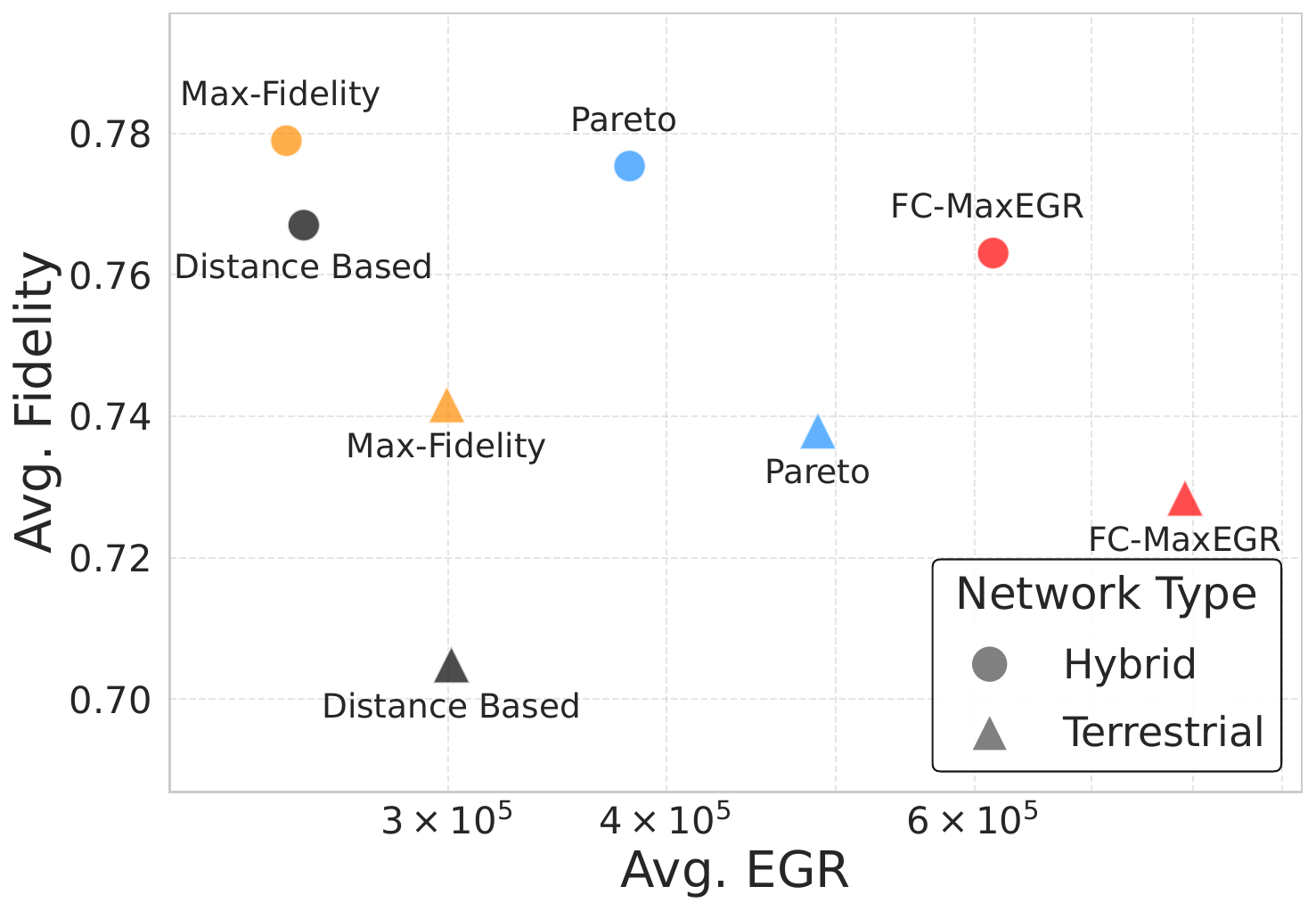}} 
    \hspace{-.3cm}
    \subfigure[]{\includegraphics[width=0.25\textwidth, height=3.2cm]{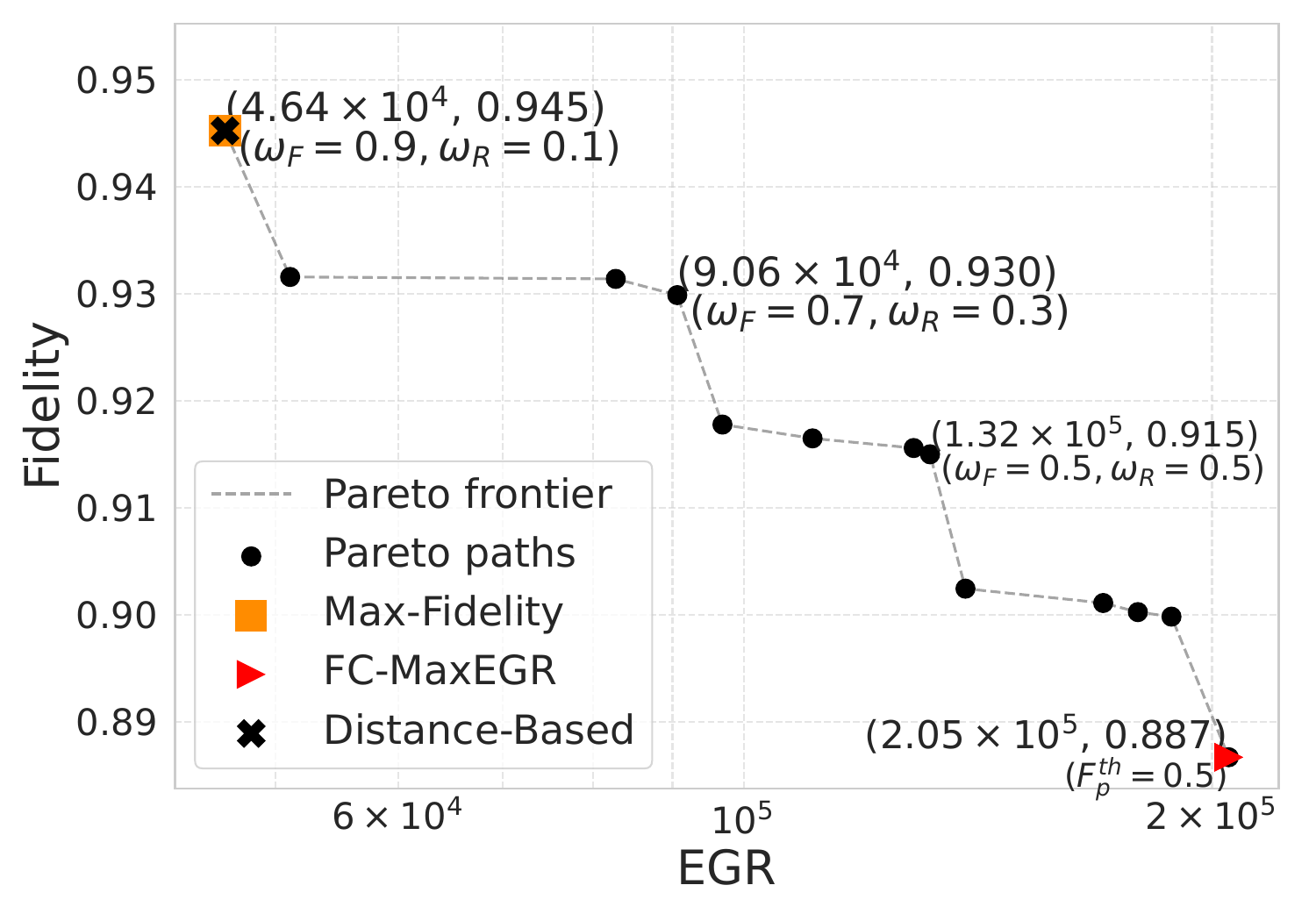}}
    \caption{Comparison of routing policies. (a) Average end-to-end EGR and fidelity. (b) A Pareto frontier for a medium-distance user pair.}
    \label{fig:routing_comp}
\end{figure}

Figure~\ref{fig:routing_comp}(a) compares the average EGR and fidelity of the routing policies over all end-user pairs. Max-Fidelity achieves the highest average fidelity but the lowest EGR, while FC-MaxEGR achieves the highest EGR with lower fidelity. Pareto Routing lies between these two extremes. Distance-Based routing achieves lower fidelity and EGR than Pareto Routing. The figure also shows the corresponding results for a terrestrial-only network. Adding satellite-assisted links increases both the average EGR and fidelity
for all routing policies.

Figure~\ref{fig:routing_comp}(b) shows a Pareto frontier for a single end-user pair at medium distance, captured during a single time slot. The Pareto paths span a wide range of EGR and fidelity values. Max-Fidelity selects the high-fidelity end of the frontier, whereas FC-MaxEGR selects a high-EGR path. The path selected by Pareto Routing depends on the weights assigned to the two objectives. Increasing $\omega_R$ moves the selected path toward higher EGR, while increasing $\omega_F$ moves it toward higher fidelity. For example, changing $(\omega_F,\omega_R)$ from $(0.9,0.1)$ to $(0.5,0.5)$ increases the selected EGR from $4.64\times10^4$ to
$1.32\times10^5$, while the fidelity decreases from $0.945$ to $0.915$. 

Thus, the Pareto frontier can be computed once. This enables the network controller to defer the final routing decision until the application's utility is known, allowing different applications to select different operating points without rerunning the routing algorithm. As demonstrated in the following subsection, this flexibility translates directly into higher aggregate application utility compared with the baseline routing policies.

\subsection{Application-Aware Routing Performance}
We next evaluate whether exposing the complete Pareto frontier translates into improved application performance. To this end, we generate a mixed workload in which each entanglement request is independently assigned one of three representative application utilities: secret key rate (SKR), negativity (NEG), or distillable entanglement (DE). For Pareto Routing (PR), the controller first computes the complete Pareto frontier and then selects the path maximizing the corresponding application utility. The remaining routing policies directly return a single path according to their routing objective.

\begin{figure}[tbp]
    \centering
    \includegraphics[width=\columnwidth]{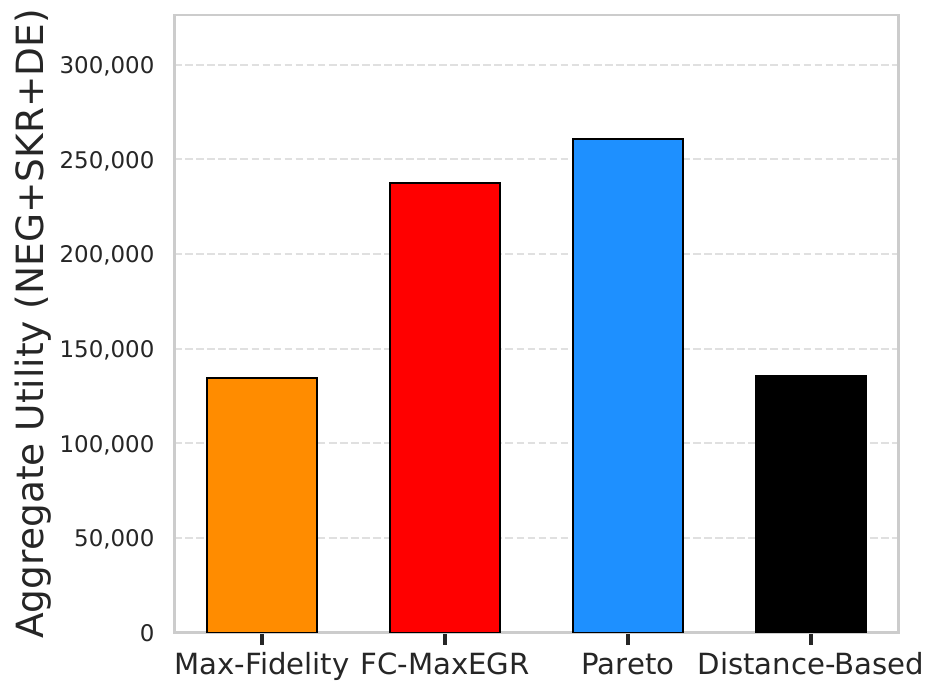} 
    \caption{Aggregate application utility achieved by different routing policies under a mixed application workload consisting of secret key rate (SKR), negativity (NEG), and distillable entanglement (DE) utilities.}
    \label{fig:pathutil}
\end{figure}

Figure~\ref{fig:pathutil} compares the aggregate application utility achieved by the routing policies over all end-user pairs. Pareto Routing achieves the highest aggregate utility,
demonstrating the benefit of exposing multiple efficient operating points. By contrast, Maximum-Fidelity routing sacrifices EGR in favor of high-fidelity paths, while Distance-Based routing often selects suboptimal routes because it relies solely on geographic separation. Among the baseline algorithms, FC-MaxEGR performs closest to Pareto Routing since it jointly considers both EGR and fidelity through a minimum fidelity constraint. Nevertheless, because FC-MaxEGR commits to a single operating point, it cannot adapt to the diverse preferences of different applications. In contrast, Pareto Routing enables each application to independently select the most appropriate operating point from the same Pareto frontier without recomputing routes.

\begin{table}[htbp]
\centering
\begin{tabular}{lcc}
\hline
Algorithm & Medium Distance & Long Distance \\
\hline
Max-Fidelity    & 838   & 2172 \\
FC-MaxEGR  & 6357  & 1228 \\
Pareto      & 3805  & 1228 \\
Distance-Based     & 0     & 1507 \\
\hline
\end{tabular}
\caption{Number of route changes for a medium-distance ($1132.63\text{ km}$) and long-distance ($3178.55\text{ km}$) pair.}
\label{tab:combined_cost}
\end{table}

\subsection{Pareto Frontier Characterization}

\begin{figure}[tbp]
    \centering
    \includegraphics[width=\columnwidth]{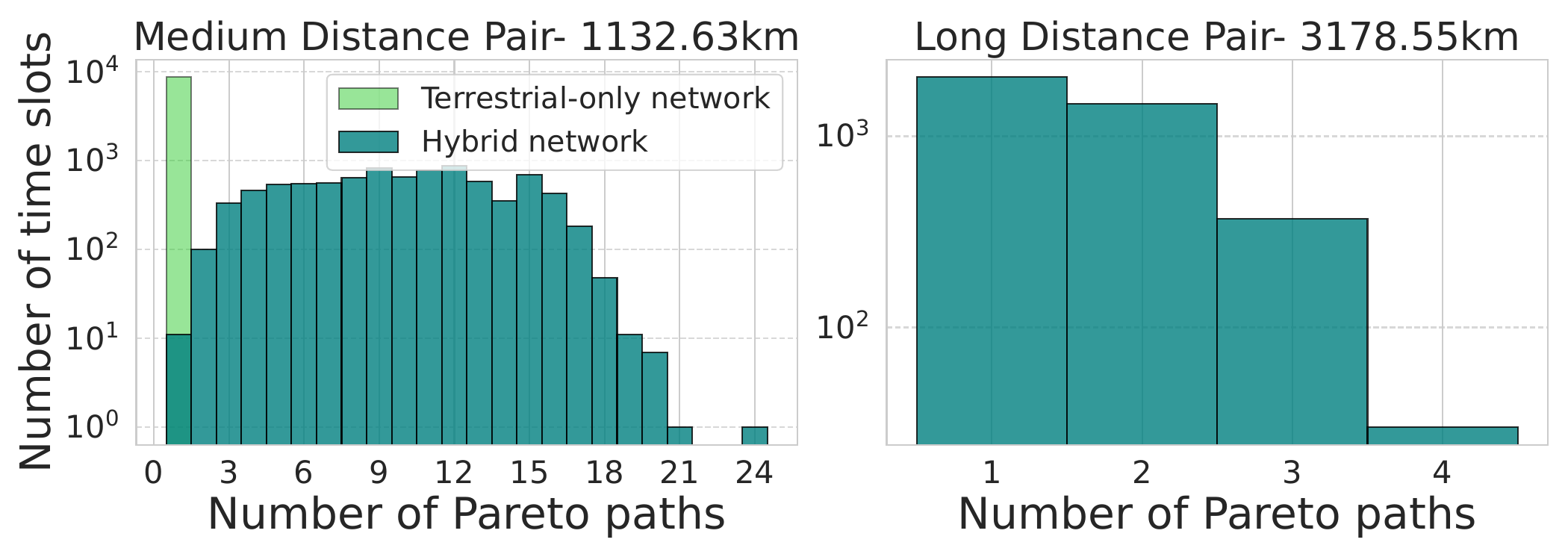} 
    \caption{Distribution of the number of Pareto-optimal paths over time for medium and long-distance end-user pairs. }
    \label{fig:pareto_path}
\end{figure}

We next examine how the hybrid network changes the number of
Pareto-optimal routing choices. Figure~\ref{fig:pareto_path} shows the distribution of the number of Pareto-optimal paths over a 24-hour period for representative medium and long-distance end-user pairs.

For the medium-distance pair, the terrestrial-only network has a single Pareto-optimal path in most time slots. In the hybrid network, the number of Pareto-optimal paths varies over time and often exceeds ten. The additional paths arise from satellite-assisted virtual links that provide different EGR--fidelity trade-offs. 

For the long-distance pair, the number of Pareto-optimal paths is smaller, with one or two paths available in most time slots. This is because fewer distinct end-to-end routes remain Pareto optimal at this distance.

These results show that satellite-assisted links can increase the number of efficient routing choices, particularly for medium-distance end-user pairs. The resulting Pareto frontier provides multiple paths from which the controller can select according to the EGR and fidelity requirements of an application.

\subsection{Route Stability}
Satellite mobility can cause routes to change over time, increasing control-plane overhead. Table~\ref{tab:combined_cost} reports the number
of route changes over 24 hours for the representative medium  and long-distance pairs. For the medium-distance pair, FC-MaxEGR changes routes most frequently, while Pareto Routing requires fewer route changes. Distance-Based does not change routes for this pair. For the long-distance pair, Pareto Routing and FC-MaxEGR have the fewest route changes, followed by Distance-Based and Max-Fidelity. These results show that considering multiple routing objectives does not necessarily lead to more frequent route changes.

\section{Conclusion}\label{conclusion}
We presented an exact bi-objective routing algorithm for hybrid
terrestrial-satellite quantum networks that jointly optimizes
entanglement generation rate (EGR) and fidelity. By exploiting the bottleneck structure of path EGR and transforming the multiplicative fidelity metric into an additive cost, we formulated routing as a MAXMIN--MINSUM bicriterion path problem. This structure enables the computation of a minimal complete Pareto set in polynomial time, providing the network controller with multiple efficient routing choices rather than committing to a single rate or fidelity optimal path. We further showed how application-specific utility functions, as well as a weighted utopia-point criterion when such utilities are unavailable, can be used to select a path from the same Pareto frontier without recomputing routes.

Our evaluation on a large-scale terrestrial network with time-varying LEO satellite connectivity shows that the hybrid architecture improves both end-to-end EGR and fidelity compared with terrestrial-only routing. In addition, satellite-assisted virtual links increase the number of Pareto-optimal paths, providing multiple operating points with different EGR-fidelity characteristics. For mixed workloads consisting of secret key rate, negativity, and distillable entanglement utilities, application-specific selection from the Pareto frontier achieves higher aggregate utility than the considered baseline routing policies. Overall, the results establish the importance of jointly considering EGR and fidelity when routing entanglement in hybrid terrestrial-satellite quantum networks.


\bibliographystyle{IEEEtran}
\bibliography{bib}

\end{document}